\documentclass[aps,pra,twocolumn,letterpaper,superscriptaddress,10pt]{revtex4-1}
\usepackage{amssymb,amsthm,amsmath,amsfonts}
\usepackage{graphicx,ulem,enumerate,mathptmx}
\usepackage[pdftex,dvipsnames,usenames]{xcolor}
\usepackage[colorlinks=true,urlcolor=blue,citecolor=blue,linkcolor=blue]{hyperref}

\begin{document}

\title{Local versus Global Two-Photon Interference in Quantum Networks}

\author{Thomas~Nitsche}
\affiliation{Applied Physics, Paderborn University, Warburger Stra\ss{}e 100, 33098 Paderborn, Germany}

\author{Syamsundar~De}\email{syamsundar.de@upb.de}
\affiliation{Applied Physics, Paderborn University, Warburger Stra\ss{}e 100, 33098 Paderborn, Germany}

\author{Sonja~Barkhofen}
\affiliation{Applied Physics, Paderborn University, Warburger Stra\ss{}e 100, 33098 Paderborn, Germany}

\author{Evan~Meyer-Scott}
\affiliation{Applied Physics, Paderborn University, Warburger Stra\ss{}e 100, 33098 Paderborn, Germany}

\author{Johannes~Tiedau}
\affiliation{Applied Physics, Paderborn University, Warburger Stra\ss{}e 100, 33098 Paderborn, Germany}

\author{Jan~Sperling}
\affiliation{Applied Physics, Paderborn University, Warburger Stra\ss{}e 100, 33098 Paderborn, Germany}

\author{Aur\'el~G\'abris}
\affiliation{Department of Physics, Faculty of Nuclear Sciences and Physical Engineering, Czech Technical University in Prague, B\v{r}ehov\'{a} 7, 115 19 Praha 1--Star\'{e} M\v{e}sto, Czech Republic}
\affiliation{Wigner Research Centre for Physics, Konkoly-Thege M. \'{u}t 29--33, H-1121 Budapest, Hungary}

\author{Igor~Jex}
\affiliation{Department of Physics, Faculty of Nuclear Sciences and Physical Engineering, Czech Technical University in Prague, B\v{r}ehov\'{a} 7, 115 19 Praha 1--Star\'{e} M\v{e}sto, Czech Republic}

\author{Christine~Silberhorn}
\affiliation{Applied Physics, Paderborn University, Warburger Stra\ss{}e 100, 33098 Paderborn, Germany}

\date{\today}

\begin{abstract}
    We devise an approach to characterizing the intricate interplay between classical and quantum interference of two-photon states in a network, which comprises multiple time-bin modes.
    By controlling the phases of delocalized single photons, we manipulate the global mode structure, resulting in distinct two-photon interference phenomena for time-bin resolved (local) and time-bucket (global) coincidence detection.
    This coherent control over the photons' mode structure allows for synthesizing two-photon interference patterns, where local measurements yield standard Hong-Ou-Mandel dips while the global two-photon visibility is governed by the overlap of the delocalized single-photon states.
    Thus, our experiment introduces a method for engineering distributed quantum interferences in networks.
\end{abstract}

\maketitle

\begin{figure*}[t]
	\includegraphics[width=\textwidth]{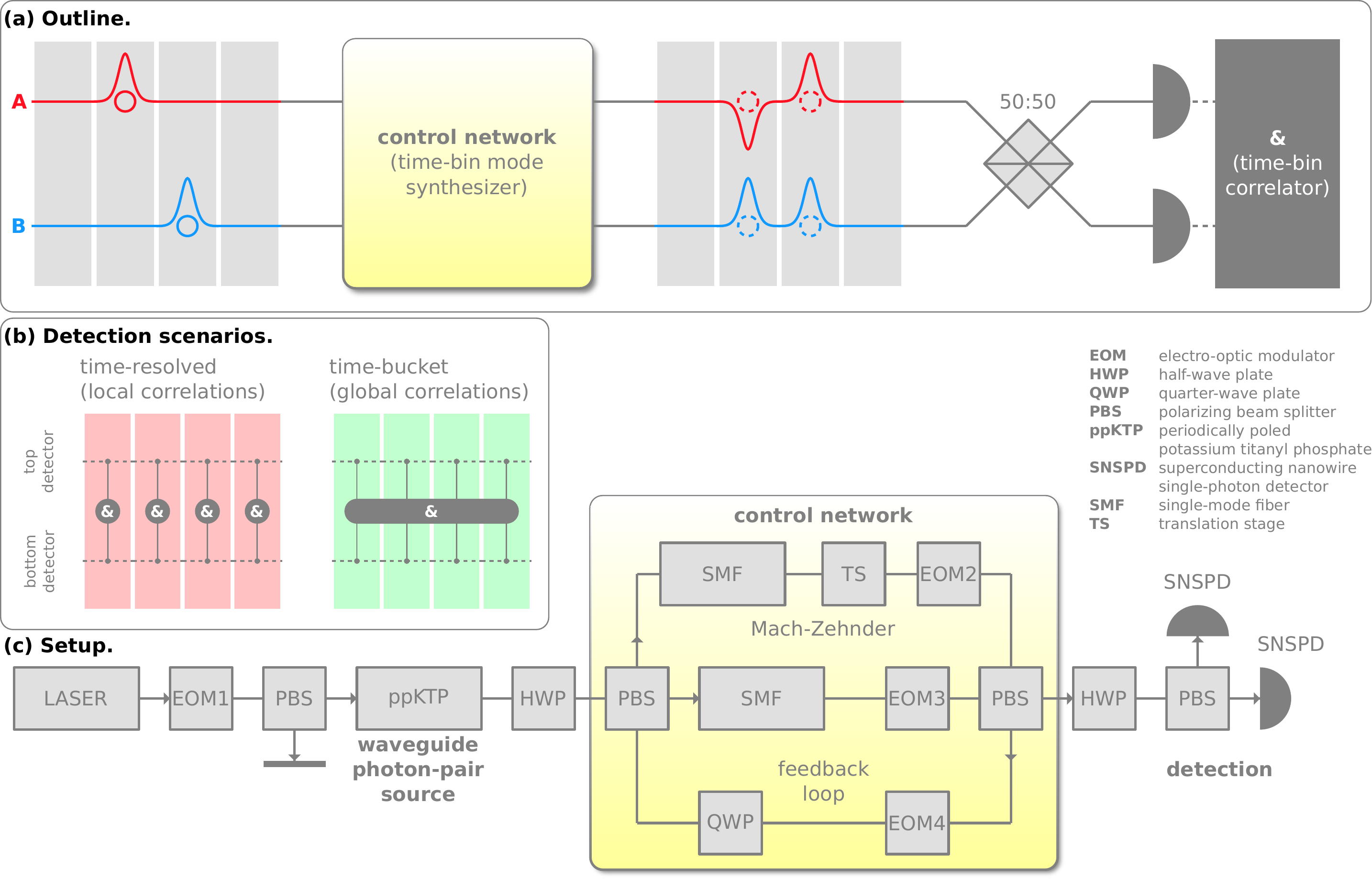}
	\caption{
		(a) Two-photon interference protocol.
		Heralded single photons $A$ and $B$ are generated in different (time-bin) modes.
		Using a highly reconfigurable network, we synthesize arbitrary mode structures over which the photons are coherently distributed.
		In a subsequent HOM-type configuration, the two photons are superimposed, and correlations are measured.
		(b) Illustration of the two fundamentally different measures of correlation.
		Local correlations involve signals from the top and bottom detectors at the same time bin, while global correlations beyond HOM interferences access coincidences across multiple time bins.
		(c) Schematics of our setup.
		Our setup comprises state-of-the-art building blocks:
		a compatible photon source, a flexible control network (implemented as a time-multiplexed Mach-Zehnder interferometer with a feedback loop and deterministic in- and out-coupling), and a versatile detection stage.
	}\label{fig:1}
\end{figure*}

\paragraph*{Introduction.---\hspace*{-2ex}}

	The naive idea that our world consists of particles, reminiscent of tiny billiard balls, which govern the laws of physics has been refuted.
	Rather, it is waves, be it classical or quantum, which describe nature best---covering areas ranging from gravity to hydrodynamics to optics to subatomic systems.
	For example, quantum field theories merely consider elementary particles as excitations of an underlying quantum field, such as photons for light \cite{Loudon73}.
	Thus, even particles must be able to interfere, which was demonstrated, e.g., in pioneering double-slit experiments with electrons \cite{Donati73}.
	To speak of genuine quantum interference, we must consider at least two particles, like in the seminal Hong-Ou-Mandel (HOM) effect in the interference of two photons \cite{HOM87}.
	With recent theoretical and technological advances in the control of quantum systems, there is a spur of interest in how such multiparticle interferences can manifest themselves in large networks \cite{Pan12,KBMW98,Metal13,TTSSGHNSW15,RFMWK16,AKJMSHRWJ17,NWXC18}.

    The control of classical coherence properties and its utilization is ubiquitous in multichannel optical systems \cite{Wiseman10}, which rely on different interfering pathways.
    Examples include coherent control in spectroscopy, chemistry, and various imaging systems.
	The strong demand for developing efficient coherent control strategies for large quantum systems arose from innovations in quantum information processing and technologies that exploit quantum coherence to its full extent \cite{KLM01,CGW13,Raussendorf01}.

	Photonic networks provide an excellent platform for studying large-scale coherence effects under designed conditions \cite{O'brien09}.
	For both practical and fundamental purposes, the quality of a network---benchmarked by stability, scalability, and reconfigurability---mainly depends on its coherence and control properties.
	Networks allow for a natural distinction of local and global features, and the introduction of multiple quantum particles to passive networks is key to many quantum communication schemes \cite{KLM01}.
	Earlier studies aimed at manipulating coherence of photons to alter the fundamental HOM effect \cite{SA88}, exposing an intricate connection of classical and quantum interference \cite{KSC92,LWHRK04,MJMTBKW17,RSG19}.
	This further inspired many studies of passive multiphoton interferences dedicated to determining or certifying nonclassicality of generated states for their potential future applications. \cite{KBMW98,Metal13,TTSSGHNSW15,RFMWK16,AKJMSHRWJ17,NWXC18}.
	Little attention, however, has been paid in multiphoton interference scenarios to the role of coherence distributed across the extent of the network.
	Yet, the inherent difference between local and global interferences suggests that their active control might offer unique insight into the interplay and convertibility of different forms of coherence within photonic networks.  

	In this Letter, we use coherent control over heralded single photons, spread over multiple nodes of a network, to demonstrate how their superposition state affects quantum interference patterns.
	We put forward correlation measures which certify the presence of two-particle quantum coherence across a linear optical network for probing local versus global coherence.
	Our implementation introduces a time-bin-multiplexing architecture with a compatible source of photons and configurable measurement for accessing various types of correlations.
	Identifying local and global correlations enables us to study contributions of the coherence properties of the source and distributed coherence properties of the network.

\paragraph*{Controlling and observing two-photon interference.---\hspace*{-2ex}}

	We outline our approach in Fig. \ref{fig:1}(a).
	The core of our system is a mode synthesizer which allows us to individually shape the time-bin mode structure of heralded single photons and, thereby, their interference characteristics.
	This is implemented using a linear photonic network, its crucial feature being reconfigurability.
	The realization of a mode synthesizer requires the ability to coherently manipulate selected degrees of freedom, such as polarization, frequency, or time-bin modes.
	Naturally, suitable single-photon sources and detectors must be available too.

	Our experiments use a time-multiplexing fiber loop setup that provides a resource-efficient, scalable, stable, and flexible platform for the implementation of networks, and which has been used, among others, to realize quantum walks \cite{ADZ93,AA11,SCPGMAJS10,NBKSSGPKJS18} and boson sampling \cite{AA11,MGDR14,Hetal17,W20}.
	Still, the dynamic operation of our network with multiple nonclassical single-photon states has not been shown to date, partly owing to the unavailability of a compatible source.
	The mode synthesis phase of our experiment is implemented by employing fast modulators and stable delay lines, resulting in each photon being coherently spread over multiple time bins.
	In the analysis phase, photons are brought to interference and measured with detectors capable of resolving individual time bins.
	In contrast to established HOM-type analysis procedures, coincidence events at the same time-bin allow for assessing local correlations, and combined coincidences across multiple time bins are used for extracting global correlations [cf. Fig. \ref{fig:1}(b)].

\paragraph*{Ideal model for local and global correlations.---\hspace*{-2ex}}

	Using the standard quantum optics formalism \cite{MW95}, we can introduce a theoretical model that describes our system in the absence of imperfections.
	We label the initial photons as $A$ and $B$, distributed in the control network over time bins as $\hat A^\dag|\mathrm{vac}\rangle=\sum_{\tau}\alpha_\tau\hat a_\tau^\dag|\mathrm{vac}\rangle$ and $\hat B^\dag|\mathrm{vac}\rangle=\sum_{\tau}\beta_\tau\hat b_\tau^\dag|\mathrm{vac}\rangle$, where $\hat a^\dag$ and $\hat b^\dag$ are bosonic creation operators for the modes under study (the index $\tau$ identifies the individual time-bin modes) and $\alpha$ and $\beta$ are the corresponding probability amplitudes.
	A superposition of these photons on a $50{:}50$ beam splitter results in output modes $(\hat a_\tau\pm\hat b_\tau)/\sqrt2$.
	The output correlations are measured with the top (${+}$) and bottom (${-}$) detector in Fig. \ref{fig:1}, represented through photon-number operators $\hat n_{\pm,\tau}$.
	This yields the first-order correlation $G_{\pm,\tau}^{(1)}=\langle\hat n_{\pm,\tau}\rangle=(|\alpha_\tau|^2+|\beta_\tau|^2)/2$, giving the same values for both detectors, and the second-order cross-correlation
	\begin{equation}
		\label{eq:G11}
		G^{(1,1)}_{\tau,\tau'}=\langle\hat n_{+,\tau}\hat n_{-,\tau'}\rangle=\frac{|\alpha_{\tau}\beta_{\tau'}-\alpha_{\tau'}\beta_{\tau}|^2}{4}.
	\end{equation}

	To arrive at a refined notion of local and global correlations, we select a set of modes $\mathbb S$, potentially being a subset of all network modes.
	For convenience, we associate the vectors $\vec\alpha=[\alpha_\tau]_{\tau\in\mathbb S}$ and $\vec\beta=[\beta_\tau]_{\tau\in\mathbb S}$ with the photons $A$ and $B$, respectively.
	This allows us to identify counts from a single detector, $G^{(1)}_{\pm}=\sum_{\tau\in\mathbb S}G_{\tau}^{(1)}=(\vec\alpha^\dag\vec\alpha+\vec\beta^\dag\vec\beta)/2$.

	Moreover, we arrive at compact formulas for correlation measures which characterize two-photon interference.
	For any selected set $\mathbb S$ of modes, we introduce local and global correlation measures via the sums
	\begin{equation}
		\label{eq:Glocal-global}
		G^{(1,1)}_\mathrm{local}=\sum_{\tau\in\mathbb S}G^{(1,1)}_{\tau,\tau}
		\quad\text{and}\quad
		G^{(1,1)}_\mathrm{global}=\sum_{\tau,\tau'\in\mathbb S}G^{(1,1)}_{\tau,\tau'},
	\end{equation}
	respectively.
	By using Eq. \eqref{eq:G11}, these correlations then obey
	\begin{equation}
		\label{eq:Glocal-global-ideal}
		G^{(1,1)}_\mathrm{local}=0
		\quad\text{and}\quad
		G^{(1,1)}_\mathrm{global}=
		\frac{
			(\vec\alpha^\dag\vec\alpha)(\vec\beta^\dag\vec\beta)
			-|\vec\alpha^\dag\vec\beta|^2
		}{2}.
	\end{equation}
	Expressions for experimentally relevant quantitites, such as normalized correlation functions, $g=G^{(1,1)}/[G^{(1)}_{+}G^{(1)}_{-}]$, and  visibilities, $V=1-2g$, can be readily obtained.

	Beyond common HOM-type correlations, our approach offers a much deeper insight into the interplay of first- and second-order coherence properties of a network by evaluating the measured coincidences and interference visibilities.
	Since local correlations $G^{(1,1)}_\mathrm{local}$ depend only on the source quality and imperfections of the network, the obtained visibilities relate to photon distinguishability at each time bin separately, exhibiting high visibility for high indistinguishability.
	In contrast, global correlations $G^{(1,1)}_\mathrm{global}$ are additionally sensitive to the synthesized mode structure by correlating coincidences over multiple time bins.

\begin{figure*}[t]
	\includegraphics[width=\textwidth]{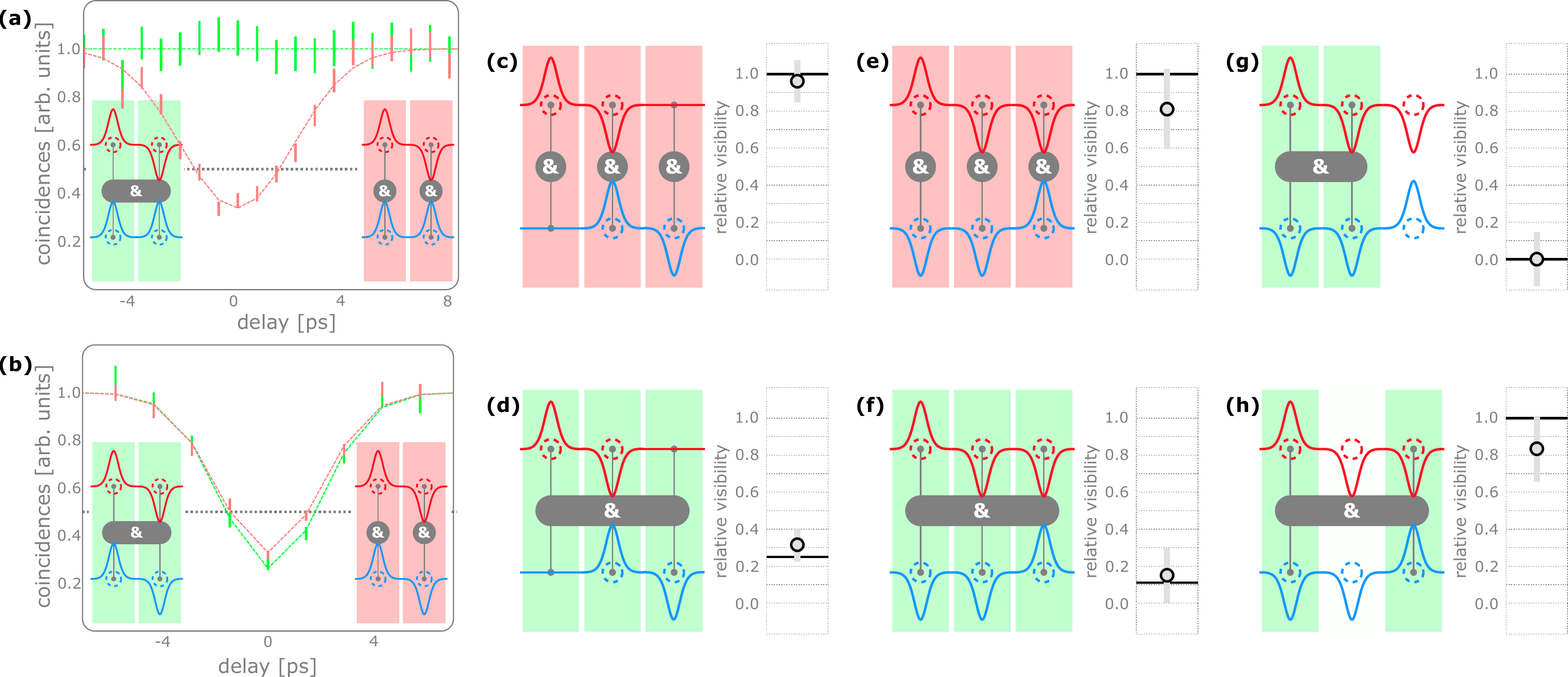}
	\caption{
		Two-photon interference patterns for various synthesized mode structures.
		Plots (a) and (b) depict the interference via coincidences, where red and green data points indicate local and global detection scenarios (see insets).
		The curves indicate values obtained from a numerical model that includes imperfections, and the dotted line marks the quantum-classical boundary, certifying photon antibunching \cite{KDM77}.
		The characteristic HOM dip can be observed via local measurements, with visibilities close to the reference (i.e., source) value; see the SM for additional information \cite{SM}.
		Global coincidences correspond to the overlap of the synthesized photon modes, showing high two-photon coherence for parallel mode structures (b) and no interference for the orthogonal case (a).
		Plots (c)--(h) depict our results for different multimode interference scenarios.
		Assuming that $V_0$ is the dominant limiting factor on all cases, the visibilities (circles including error bars) are normalized by this value.
		The good agreement with the ideal model (thick solid line) justifies this assumption.
	}\label{fig:2}
\end{figure*}

\paragraph*{Implementation.---\hspace*{-2ex}}

	At the core of our experimental setup lies a fiber-based unbalanced Mach-Zehnder interferometer with a feedback loop, as outlined in Fig. \ref{fig:1}(c), that serves as a dynamically reconfigurable, time-multiplexing network \cite{Nitsche16}.
	The length difference of the single-mode fibers at the two interferometer arms sets the time-bin spacing (${\sim}105\,\mathrm{ns}$).
	A translation stage (TS) allows for a fine scanning of the time delays in the picosecond regime between the two interfering photons.
	The network contains fast electro-optic modulators (EOMs), capable of implementing controlled polarization rotations at any time bin.
	EOM2 and EOM3 ensure deterministic in- and out-coupling of the photons, whereas EOM4 allows synthetization of complex mode structures by programming appropriate switching patterns. 
	A detailed description of our time-multiplexing scheme along with its specifications, and the details of our switching patterns, including the actual timings of the EOM operations for synthesizing the desired photon modes, are provided in the Supplemental Material (SM) \cite{SM}.  

	In addition, we implement a new source via a type-II parametric down-conversion process in periodically poled potassium titanyl phosphate waveguide as an engineered source of heralded single photons with high spatial and spectral purity \cite{HABDMS13}.
	A picosecond pump laser at $775\,\mathrm{nm}$ and a bandwidth of ${\sim}0.3\,\mathrm{nm}$ together with a $2.5\,\mathrm{cm}$ long waveguide generates relatively broad (${\sim}2.7\,\mathrm{ps}$) photon pulses at telecom wavelength.
	These picosecond photon pulses barely suffer from the difference in dispersive broadening in the fibers, thus maintaining good indistinguishability even after several round-trips through the network.
	To ensure that the two interfering photons are generated in desired time bins, we implement pulse picking on the pump laser using EOM1 and a polarization beam splitter (PBS).
	As a measure of the source quality, we obtain a visibility of the HOM coincidence count suppression of up to $V_0=0.801\pm 0.067$, limited by the residual spectral distinguishability as well as higher photon-number terms in the heralded photon states. 
	It is worth emphasizing that the nonunit visibility of our source is not an obstacle to accomplish our main objective of characterizing local and nonlocal coherence properties.
	To this goal, however, it is imperative to monitor the source visibility in each experiment that serves as a reference to exclude the source imperfections from the interference of the synthesized modes (see the SM \cite{SM}). 

	Our detection scheme consists of two superconducting nanowire single-photon detectors with dead time and jitter well below the time-bin spacing, together with a PBS for separating the two polarizations, thus allowing both polarization and time-bin resolved measurements.

\paragraph*{Results.---\hspace*{-2ex}}

	In Fig. \ref{fig:2}, we present the results of our instructive two-photon interference analysis.
	In Figs. \ref{fig:2}(a) and \ref{fig:2}(b), the measured local and global coincidences are plotted against the delay introduced by the TS.
	Figure \ref{fig:2}(a) corresponds to the case when the photons are distributed over two time bins such that their mode structures are expected to be orthogonal, described by two orthogonal vectors $\vec{\alpha}$ and $\vec{\beta}$, while correlations obtained for photons that are designed to have identical mode structures are given in Fig. \ref{fig:2}(b).
	In both cases, the local correlations behave identically, and with visibilities significantly exceeding the classical threshold, certify local quantum features.
	Moreover, as expected from our simulations, the local visibilities resemble the corresponding source visibility.
	Global correlations that incorporate network effect via modal superposition, however, exhibit a remarkably different behavior depending on the global mode structure of the photons.
	For the first case [Fig. \ref{fig:2}(a)], no interference is observed, certifying vanishing mode overlap, i.e., orthogonality of the photons' global mode structures $\hat A$ and $\hat B$.
	However, for the latter [Fig. \ref{fig:2}(b)], the visibility of interference is roughly equal to those of local correlations, indicating an almost perfect mode overlap, thus validating that the photons' modes are identical.

    The complete disappearance of the HOM dip in the orthogonal case in contrast to its maximum visibility in the parallel case (only limited by the source quality) highlights the near-perfect performance of the network and control over relative phases between the photons.
    Since the imperfection of the source can be factored out from the global visibilities, the markedly different behavior of global and local correlations allows for exact quantification of the nonlocal coherence, linked to the network and the local coherence dependent on the source.
    It is important to note that the certification of nonclassicality via a visibility $>50\%$ is only relevant for source (i.e., local) coherence, which is successfully achieved at this point.
    Studying the global coherence shows that our nonclassical states can exhibit completely different two-photon interference behavior with visibilities which we can engineer to our will by designing appropriate global mode overlaps.
    For instance, the global visibility in Fig. \ref{fig:2}(a) is zero although the states are clearly nonclassical single-photon states (as certified by local interference).

	To explore the impact of coherent control for more complex mode structures, we synthesized various single-photon states involving three time bins, Figs. \ref{fig:2}(c)--\ref{fig:2}(h).
	To factor out the effect of initial impurities, as argued above, we normalize all obtained visibilities by the corresponding reference (source) visibility and observe remarkably good agreement with the ideal model (thick solid lines). 
	For additional details about the source visibility and the normalization, see the SM \cite{SM}.

	For Figs. \ref{fig:2}(c) and \ref{fig:2}(d), we considered photons spreading across two time bins but with a relative offset of one time bin.
	This is achieved through our unique control over the parameters $\vec \alpha$ and $\vec \beta$.
	This does not change the maximal local interference, Fig. \ref{fig:2}(c).
	However, due to the reduced overlap between the photon states, our model predicts a reduced global visibility of $25\%$ in Fig. \ref{fig:2}(d) to which our experimental data agree with within the error margin.
	(Errors are obtained from standard statistical error analysis and error propagation.)

	In Figs. \ref{fig:2}(e)--\ref{fig:2}(h), we depict our results with photons in superposition of three modes.
	The preservation of the quality and the quantum nature of two-photon interference through the network is certified by local correlations shown in Fig. \ref{fig:2}(e).
	Cross-correlations between all pairs of time bins, Fig. \ref{fig:2}(f), are expected to yield a global visibility of $1/9\approx 11\%$ for the generated pulse shapes, again being confirmed by our data.
	To reveal coherence between parts of the network, we consider correlations restricted to subsets of modes $\mathbb S$.
	For instance, when restricted to the first two time bins, the two photons have orthogonal submode structures [Fig. \ref{fig:2}(g)], thus yielding no visible interference.
	However, coincidences from time bins one and three [Fig. \ref{fig:2}(h)], in which the photons have identical submode structure (up to a global phase), exhibit quantum coherence limited only by the photons' distinguishability [compare to Fig. \ref{fig:2}(e)].
	Additional details in the SM \cite{SM} further support the excellent performance of our network.

	Therefore, these results demonstrate how classical coherent control over the mode synthesizing network can be used to alter global quantum interference across several optical modes.
	Thus, by tailoring the time-bin-distributed shape of the input quantum light, we can generate and analyze the intricate details of coherent correlations of interfering quantum particles, despite our source not producing perfect single-photon states.

\paragraph*{Summary and conclusion.---\hspace*{-2ex}}

	In summary, we established a generic scheme for jointly controlling and characterizing local and global coherence effects in the interference of multiple quantum particles.
	Using a time-multiplexed network for our on-demand mode synthesis, we determined interference visibilities to quantify the amount and kind of quantum coherence imprinted in the temporal distribution of two photons.
	Thereby, our experiment demonstrates an intricate interplay between classical mode interference and quantum coherence, and it serves as a testimony for how classical coherence can be used to govern quantum effects.
	Beyond purely assessing standard quantum HOM interference, our framework applies to any network architecture for the realization of a manifold of multimode coherence phenomena at will.
	
	In the context of time multiplexing, our concepts can be intuitively related to extending standard HOM interference.
	However, this framework is applicable to any network implementation, e.g. spatial or frequency multiplexed, yielding a powerful tool for realizing and analyzing networkwide quantum coherence phenomena.
	The advantage of our characterization scheme is that instead of complex measurement settings and elaborate communication protocols between distant nodes of the network, it relies only on sufficient synchronization that allows individual parties to identify different runs, hence coincidence events.
	In addition to the high-performance network and intricate detection part, a compatible source of single photon has been implemented, requiring a setup overhaul and optimization to reliably operate in the single-photon regime instead of using bright coherent light.

	Our results certify an unprecedented level of control that extends over multiple modes and which enables us to manipulate quantumness not only locally, but globally.
	This includes engineering interference between arbitrarily selected parts of the full system.
	Our coherent control renders it possible to purposefully alter our system toward any desired interference for studying the rich landscape of quantum superpositions of particles.
	Furthermore, our network has no fundamental restrictions regarding future increments of the number of modes and photons, thus paving the route for photonic quantum information science, such as quantum simulators \cite{Guzik12} and remote state preparation protocols \cite{Bennett2005}, which exploit different forms of multiphoton quantum interference phenomena.

\paragraph*{Acknowledgments.---\hspace*{-2ex}}

	The Integrated Quantum Optics group acknowledges financial support through the European Commission through the ERC project QuPoPCoRN (Grant No. 725366) and the Gottfried Wilhelm Leibniz-Preis (Grant No. SI1115/3-1).
	A. G. and I. J. received financial support by the Czech Science foundation (GA\v{C}R) Project No. 17-00844S and by RVO 14000.
	I. J. acknowledges funding from the project ``Centre for Advanced Applied Sciences,'' Registry No. CZ.02.1.01/0.0/0.0/16\_019/0000778, supported by the Operational Programme Research, Development and Education, co-financed by the European Structural and Investment Funds and the state budget of the Czech Republic,
	and A. G. by the National Research Development and Innovation Office of Hungary under Project No. K124351.

%%%%%%%%%%%%%%%%%%%%%%%%%%%%%%%%%%%%%%%%%%%%%%
%%%%%%%%%%%%%%%%%%%%%%%%%%%%%%%%%%%%%%%%%%%%%%
%%%%%%%%%%%%%%%%%%%%%%%%%%%%%%%%%%%%%%%%%%%%%%
\onecolumngrid
\section*{Supplemental Material}
\twocolumngrid

%%%%%%%%%%%%%%%%%%%%%%%%%%%%%%%%%%%%%%%%%%%%%%%%%%%%%%%%%%%%%%%%%%%%%%%%%%%%%
%%%%%%%%%%%%%%%%%%%%%%%%%%%%%%%%%%%%%%%%%%%%%%%%%%%%%%%%%%%%%%%%%%%%%%%%%%%%%

\section{Detailed description of the setup}

\begin{figure*}
	\centering 
	\includegraphics[width=\textwidth]{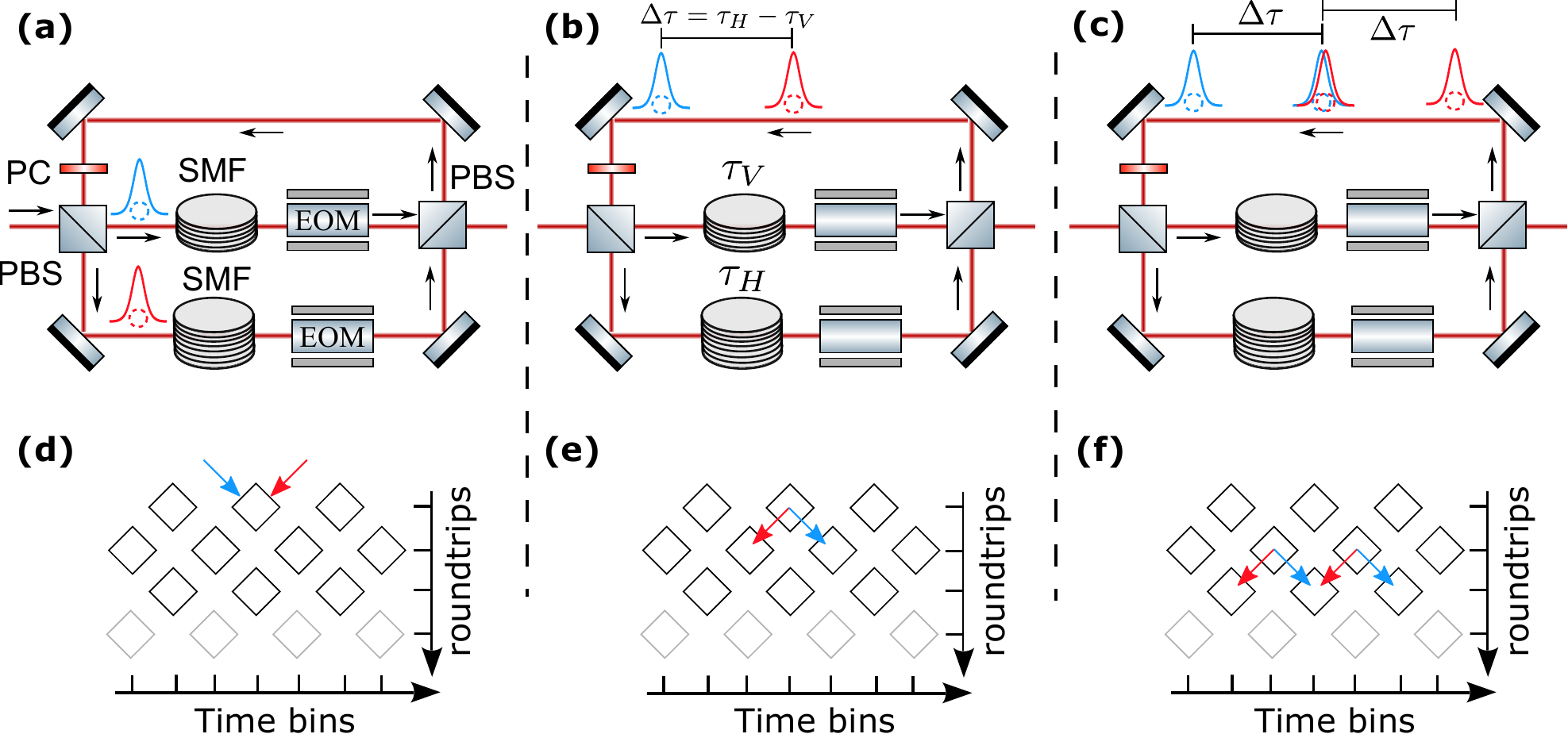}
	\caption{
	    Illustration of how time-multiplexing is implemented in our setup. 
	    (a): Incident light undergoes a polarisation-dependent splitting at a polarising beam splitter (PBS). 
    	Horizontal components are represented by red pulses, vertical light by blue pulses. 
    	(b): In the two single-mode fiber (SMF) based arms, the pulses are delayed by $\tau_H$ and $\tau_V$, introducing a delay between them, i.e., a time-bin spacing of $\Delta \tau = \tau_H - \tau_V$.
    	The two paths are merged again at the second PBS, mapping the outcome of the initial splitting operation into the time domain.
    	The pulses are fed back to first PBS.
    	Before they are split here again, they can undergo, for instance, a polarization rotation via the polarization controller (PC). 
    	(c): By repeating the splitting and delaying in time, we achieve the desired spreading of the photon's wavefunction with increasing round-trips over an increasing number of time bins. 
    	This spreads lends itself to the pyramidal representation which is found in (d)--(f) and will be used throughout this Supplemental Material for illustration purposes.
	}
	\label{fig:Illustrate_time_multi}
\end{figure*}

    Here we present the details of our time-multiplexing network that has proven to be resource-efficient, highly stable and homogeneous, scalable, and dynamically reconfigurable \cite{SCPGMAJS10,Nitsche16}. 
    To start with, we briefly recall the underlying concept of our time-multiplexing scheme  \cite{SCPGMAJS10}.
    The principle of time-multiplexing using a fiber-based Mach-Zehnder interferometer is illustrated in Figure \ref{fig:Illustrate_time_multi}(a)--(c).
    This includes  polarization dependent splitting at the first polarizing beam splitter (PBS), propagation of horizontal ($H$) and vertical ($V$) polarization through long and short fibers, respectively, and, finally, a coherent recombination of the two paths at the second PBS to introduce a well-defined delay ($\Delta \tau = \tau_H - \tau_V$) between the polarizations. 
    Note that if the EOMs are switched off, then both the polarizations exit from the right port of the second PBS without getting coupled into the network;
    if the EOMs are set such that they switch the respective polarization, then the light is redirected to the free-space feedback loop, which routes the light back to the first PBS for the next step. 
	Afterwards, each of the time bins is split up again, resulting in a cascaded process; see Figure \ref{fig:Illustrate_time_multi}(d)--(f). 
	In totality, this leads to the controlled distribution of the input photon wavepacket into multiple time bins.
	Finally, both polarizations at each time bin are again switched with the EOMs to out-couple them from the network and redirect towards the detection unit. 
    
	The time-multiplexing network of our setup, indicated by the yellow-shaded region of Fig. 1(c) of the main text, contains two fibers of length $\sim473\,\mathrm{m}$ and $\sim453\,\mathrm{m}$, leading to a time-bin separation of $\Delta \tau \sim105\,\mathrm{ns}$ and a roundtrip time of $\sim2.3\,\mathrm{\mu s}$ through the loop. 
	This time-bin spacing allows us for the measurement of time-bin-resolved coincidence counts using a highly efficiency ($>90\%$) and low jitter ($\sim50\,\mathrm{ps}$) superconducting nanowire single-photon detectors (SNSPDs), having a deadtime of $\sim100\,\mathrm{ns}$. 
	Moreover, the roundtrip time is chosen such that the photon wavepacket can be distributed to adequately large number of time bins (up to $20$), guaranteeing no time-bin interlacing. 
	It is also worth mentioning that all our EOMs are sufficiently fast with a switching time of ${\sim}10\,\mathrm{ns}$, thus enabling us addressing individual time bins. 
	The SNSPD outputs are recorded using a Swabian time tagger which has a low timing jitter ($\sim20\,\mathrm{ps}$), facilitating the time-resolved coincidence measurement. 
	A similar time-multiplexing scheme has been used for the demonstration of diverse quantum walk experiments using coherent laser pulse as the input state.

	Here we significantly extend the capability of the previous setup by the inclusion of single-photon input states, paving the route for our multiphoton interference experiment. 
	This experimental leap is accomplished on one hand by significantly improving the stability and the efficiency of the setup and, on the other hand, by developing a single-photon source that is compatible with the network. 
	Firstly, for a single-input quantum walk experiment with coherent light the interfering beams propagate through the same fibers albeit in a different sequence, thus the relative temporal drift between the interfering paths is canceled. 
	However, for the current experiment, the two interfering photons propagate through somewhat different optical paths, thus obtaining a good interference visibility demands minimization of temporal drifts.
	An in-depth characterization of our network stability shows that temperature fluctuations are the main detrimental factor.
	Therefore, the fiber-based interferometer arms are temperature stabilized to reduce the temporal drifts of the photon pulses below a picosecond level.
	Secondly, a low loop efficiency, for instance, in quantum walk experiments with coherent laser light was compensated by straightforwardly increasing the input laser power. 
	However, for single-photon inputs utilized here, increasing the generation rate by increasing the pump power is not a viable option since the increased multiphoton components at high pump power deteriorates the HOM visibility (see next section).

	Moreover, the success rate in a multiphoton interference experiment scales exponentially with efficiency, with the photon numbers being at the exponent.
	For example, the present experiment involves four-fold coincidence measurement which depends on the fourth power of loop efficiency, thus requires a high-loop efficiency for obtaining sufficient counts in reasonable times.
	The current efficiency is increased to $> 80\%$ by upgrading the network to operate at telecom wavelength as well as incorporating active, deterministic in- and out-coupling.
	This led to an overall Klyshko-efficiency of $\sim 30\%$ that takes into account several effects (transmission in the waveguide, $\sim 75\%$; coupling from the source to the network, $\sim 75\%$; loop efficiency, $\sim 80\%$; coupling from the network to the detection unit, $\sim 75\%$; and detector efficiency, $\sim 90\%$).
	Overall, this ensures that the total data collection time stays within the stability range of our experiment, $\sim 12$ hours, for instance, in a four-fold coincidence detection that includes five round-trips through the network (cf. Fig. \ref{fig:Switching_pattern_three-bin_tau0}).
	%In the following, we discuss the details of the compatible single-photon source along with its in-depth characterization. 
	
% 	\begin{itemize}
% 	    \item synchronization of EOMs
% 	    \item EOM switching matrices
% 	    \item: trigger rate 380 KHz, each measurement 2000 seconds 
% 	    \item: coincidence errors: Poissonian error.
% 	\end{itemize}
	
\section{Source characterization}

\begin{figure*}
    \centering
    \includegraphics[width=\textwidth]{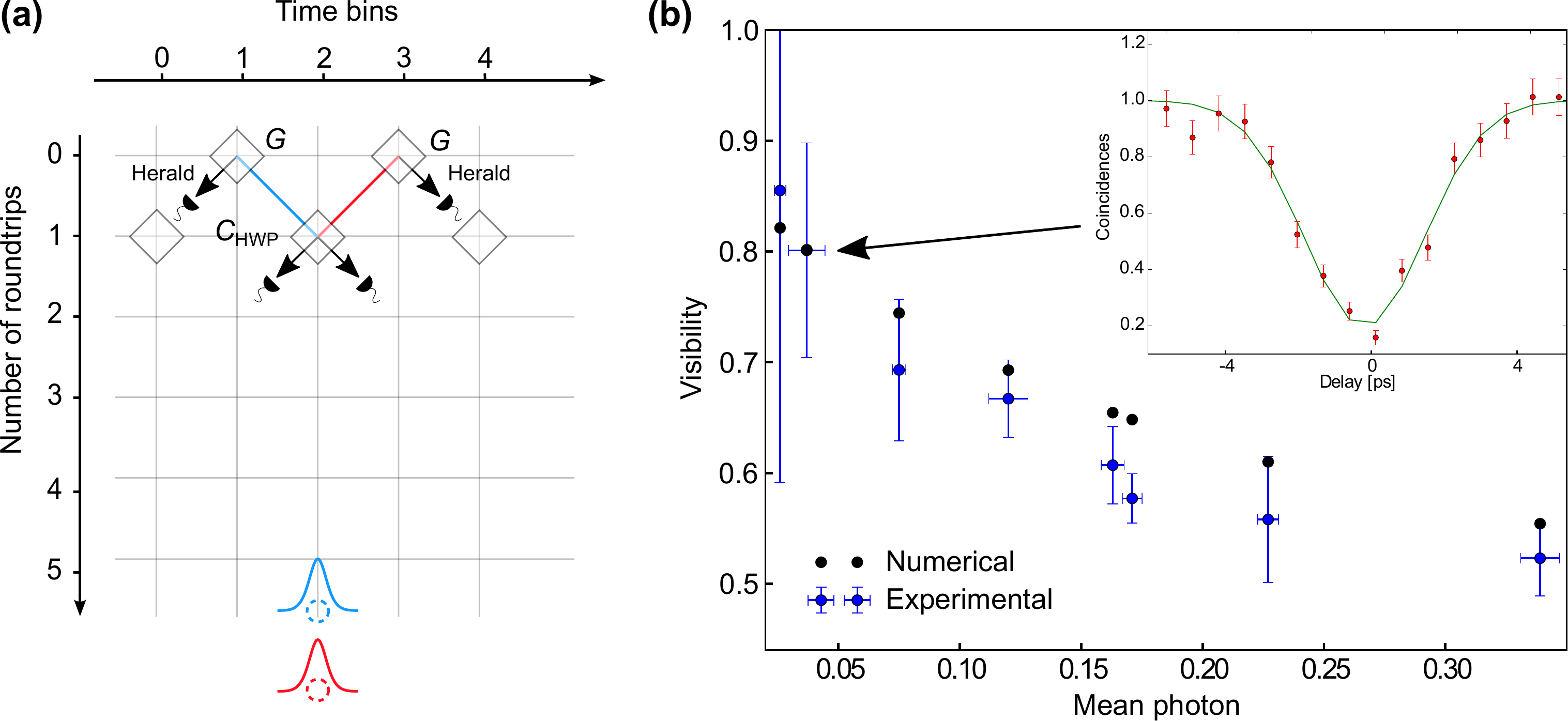}
    \caption{
        (a): The simplest switching patterns that is implemented for checking the quality of our time-multiplexed single-photon source. 
        (b): Variation of the HOM visibility with the mean photon to estimate the multiphoton effect.
        Blue dots stand for the experimentally obtained values with the error bars on the visibility (mean squared deviation of the measured data points, red dots, from the fit, green line, in the inset) and the error bars on the mean-photon number $\bar{n}$ that considers variation of $\bar{n}$ within a measurement interval mainly due to the fluctuations of the lab temperature.
        Black dots indicate the expected visibilities from our rigorous numerical model showing good matching with the experiment.
        Inset: Measured coincidences (red dots) for a mean-photon number $\bar{n}=0.0165\pm 0.0067$ that lead to a visibility $V_0=0.802\pm 0.067$.
    }
    \label{fig:Vis-Source_vs_mean-photon}
\end{figure*}

    % \begin{itemize}
    %     \item ps source: minimize the effect of dispersion in the fibers that deteriorates the visibility 
    %     \item time-multiplexing: down-sampling of pump pulse (13 ns) to fit the time-bin spacing of 105 ns of the network: every 8th pulse is picked. 
    %     \item Klyshko efficiency: 30\%, HOM visibility varies with generation rate. 
    %     \item During the experiment the generation rate is chosen around 0.15, which is good trade-off between a good HOM visibility and experimental stability. 
    % \end{itemize}

    We realized a new parametric down-conversion (PDC) based single-photon source that is compatible with our time-multiplexed network. 
    The key compatibility requirements are (i) telecom wavelength single-photons for minimizing loss in the fiber network, (ii) picosecond photon pulses to mitigate dispersion-induced differential temporal broadening of the two mode-synthesized interfering photons, and (iii) controlled time-multiplexed generation of the photons at desired time bins of the network. 
    The detail of the course design is schematized in Fig. 1(c) in the main text (at the left of the yellow shaded region).

    We achieve the above criteria by developing a source based on a $2.5\,\mathrm{cm}$ long ppKTP waveguide (from AdVR) which is pumped with a Ti:Saphire laser (MIRA) delivering picosecond pulses centered at $775\,\mathrm{nm}$ with a bandwidth of $0.3\,\mathrm{nm}$.
    To synchronize photon-generation events with the time bins of our network, the repetition of the pump pulses are down-sampled from $\sim 13\,\textrm{ns}$ to modulo $\sim 105\,\textrm{ns}$ (time-bin spacing) via pulse-picking using an EOM (EOM1) in combination with a PBS (cf. Fig. 1 in the main text).
    For example, notice the switching pattern used for the characterization of the source where the effects of multiple round-trips through the network are absent (Fig. \ref{fig:Vis-Source_vs_mean-photon}).
    The two photon generation events (G) are separated by two time bins ($2\Delta\tau\sim 210\,\textrm{ns}$) such that the $V$-photon (blue) form the first generation event propagating through the short path meets the $H$-photon (red) from the second generation event that traverses through the long path.
    These two photons are then redirected towards to the detection unit using the in- and out-coupling EOMs.
    The other photons (black) from the two generation events, that act as heralds for the interfering photons, are also routed towards the detection unit.
    The detection units contains two SNSPDs, followed by a half-wave plate (HWP) and a PBS which allows for the interference of the $V$- and the $H$-photon. 
    Finally, we detect the two PBS outputs along with the two heralds, i.e., four-fold coincidences, for the analysis of HOM interference.

    To benchmark our source, we measured the HOM visibility by varying the mean photon number $\bar{n}$ as shown in Fig. \ref{fig:Vis-Source_vs_mean-photon}(b). 
    This shows how the HOM visibility drops for high $\bar{n}$ as a result of the enhanced contribution from the unavoidable higher order photon-number components in a PDC-based source.
    Moreover, we find that even for very low  $\bar{n}$ the visibility does not approach to unity.
    Our careful analysis links this to several imperfections: residual temperature-induced temporal drifts (estimated visibility $0.98$), slight non-degeneracy of the generated photons due to the limited tunability of the Ti:Sa pump working in the picosecond regime (estimated visibility $0.95$), and the existence of slightly multimode feature of the source in Schmidt-mode basis (estimated visibility $0.90$) due to the imperfect filtering of the phase-matching side lobes.
    These effects contribute to the overall visibility, limiting it to a maximum value of $\sim 0.83$.

    A numerical model considering above imperfections along with various noises (unfiltered pump, unwanted type-0 PDC process) explains the measured visibilities; see Fig. \ref{fig:Vis-Source_vs_mean-photon}(b). 
    For the main results of this work, we choose the values between $\bar{n}=0.1$ and $\bar n=0.15$ to record sufficient number of counts within a reasonable time although this restricts the reference (source) visibility from $V_0=0.7$ to $V_0=0.6$.
    It is important to mention here that this limited source visibility does not hinder us to distinguish the local (only source effect) and the global (both source and network effect) two-photon interference phenomena which is the main motivation of this work. 

\section{Network characterization}

    In this section, we provide additional information, further supporting the main results of this work. 
    In particular, we illustrate the timings of various transformations (transmission, $T$; reflection, $R$; equal mixing of the two polarizations with a quarter-wave plate $C_\textrm{QWP}$ and a half-wave plate $C_\textrm{HWP}$) by means of the switching patterns in addition to the relevant results of four-fold coincidence measurements.
    In all the following cases, the switching patterns and the coincidence detection are triggered at a rate of $\sim 63\,\mathrm{kHz}$ with a trigger signal derived from the pump laser. 
    Each measured data point in the coincidence plots (red dots) are recorded for $2\,000\,\mathrm{s}$, and the error bars there correspond to the statistical error. 
    The green line indicates the numerical fitting, and the error in the HOM dip visibility is calculated from the mean squared deviation of the measured data points from the fit.   

	In the mode-synthesizing network (yellow shaded region), we are interested in the implementation of mainly three transformations ($T$, $R$, and $C_\textrm{QWP}$) at the desired time bins to construct the envisaged photon temporal modes. 
	This time-bin dependent polarization rotation is achieved via the introduction of a fast-swithing EOM ($\textrm{EOM}_4$) followed by a quarter-wave plate (QWP) in the feedback path.
	The action of a QWP aligned at an angle $45^\circ$ with respect to the polarization basis $\{H,V\}$ reads as 
	\begin{equation}
		C_{\textrm{QWP}}=\frac{1}{\sqrt{2}}
		\begin{pmatrix}
			1 & -i \\
			-i & 1
		\end{pmatrix}.
	\end{equation}
	The EOM operation can be written as
	\begin{equation}
		C_{\textrm{EOM}}=
		\begin{pmatrix}
			\cos{\phi} & -i\,\sin{\phi} \\
			-i\,\sin{\phi} & \cos{\phi}
		\end{pmatrix},
	\end{equation}
	where the phase $\phi$ can be tuned by varying the voltage applied to the EOM.
	Their combination leads to the transformation
	\begin{equation}
		C_{\textrm{EOM}}\, C_{\textrm{QWP}}=
		\begin{pmatrix}
			\cos{\theta} & -i\,\sin{\theta} \\
			-i\,\sin{\theta} & \cos{\theta}
		\end{pmatrix},
	\end{equation}
	using $\theta=\phi+\frac{\pi}{4}$.
	%and the identities $(\cos{\phi} - \sin{\phi})/\sqrt 2 = \cos{\theta}$ and $(\cos{\phi} + \sin{\phi})/\sqrt 2 = \sin{\theta}$. 
    The EOMs allow three different voltage settings, $v\in\{-v_1,0,+ v_1\}$, during a single experimental run.
	In particular, $v=0$ corresponds to $\phi = 0$, leading to transformation $C_\textrm{QWP}$ that equally mixes $H$- and $V$-photons.
	We chose $v=\pm v_1$ such that $\phi=\mp\pi/4$.
	This yields transformation $T$ that leaves the polarization states unchanged and $R$ that switches the polarizations.
	Notably, we find that these three transformations are sufficient to engineer the photon temporal modes according to our requirements as demonstrated below.
	The EOM operations are controlled by designing appropriate voltage-switching patterns and careful synchronizations of the EOM switchings and the arrival of the photon pulses. 
	
	In the following Figs. \ref{fig:Switching_pattern_two-bin_tau0}, \ref{fig:Switching_pattern_two-bin_tau-1}, and \ref{fig:Switching_pattern_three-bin_tau0}, we show additional data to support the results presented in Fig. 2 in the main text. 

% \begin{thebibliography}{99}
% 	\bibitem{SCPGMAJS10}
% 		A. Schreiber, K. N. Cassemiro, V. Poto\v{c}ek, A. G\'{a}bris, P. J. Mosley, E. Andersson, I. Jex, and C. Silberhorn,
% 		Photons Walking the Line: A Quantum Walk with Adjustable Coin Operations,
% 		Phys. Rev. Lett. \textbf{104}, 050502 (2010).
% 	\bibitem{Nitsche16}
% 	    T. Nitsche, F. Elster, J. Novotn\'{y}, A. G\'{a}bris, I. Jex, S. Barkhofen, and C. Silberhorn, 
% 	    Quantum walks with dynamical control: graph engineering, initial state preparation and state transfer,
% 	    New. J. Phys. \textbf{18}, 063017 (2016).
% \end{thebibliography}

\begin{figure*}
    \centering
    \includegraphics[width=\textwidth]{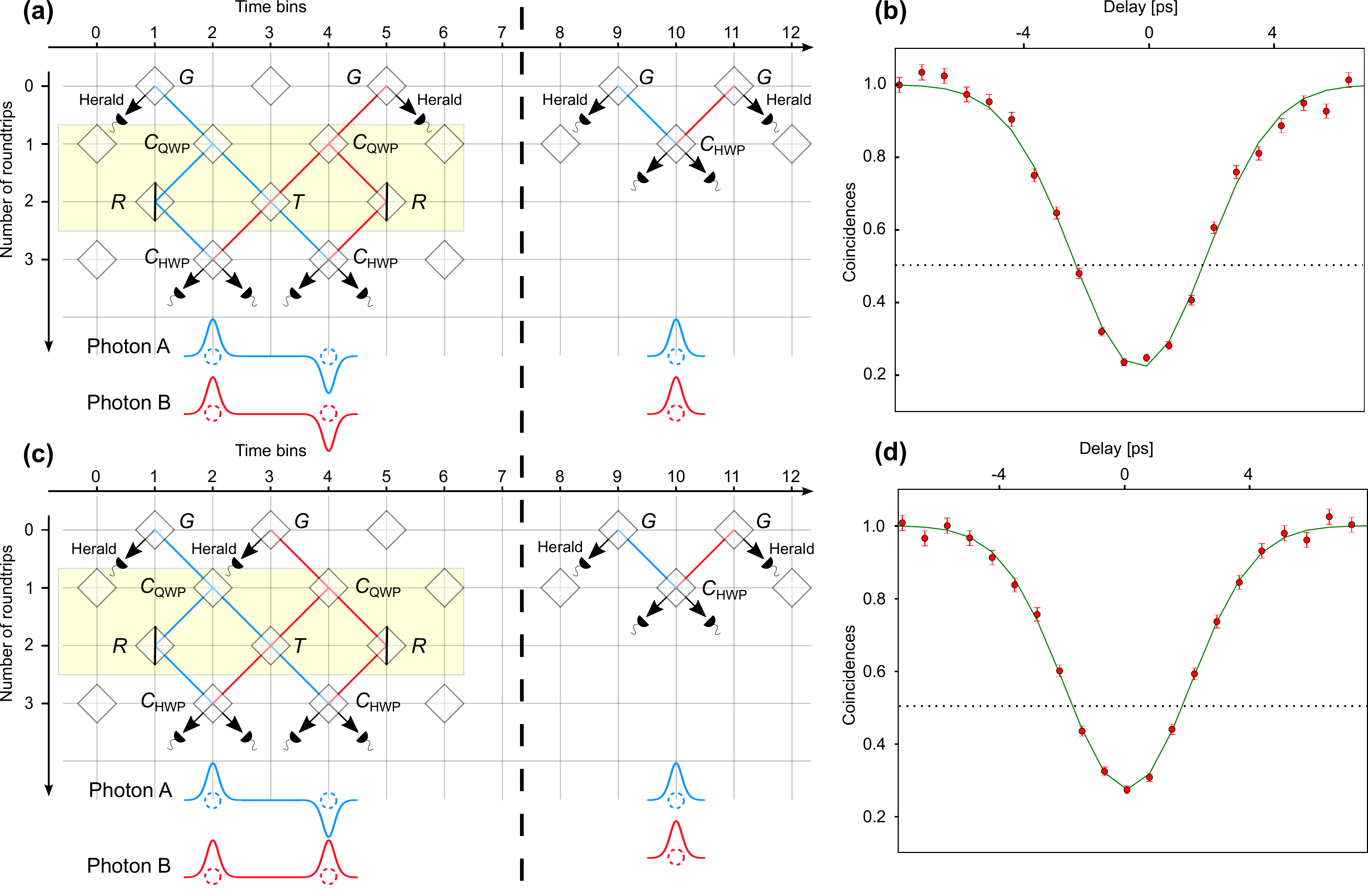}
    \caption{
        Here we present addition information in support of the results presented in Figs. 2(a) and (b) of the main text.
        (a) and (c): The switching patterns implemented for the preparation of the parallel and orthogonal temporal modes.
        On the left of the dashed line, we present the actual operations and their timings for the generation (top), mode-synthesising (yellow shaded region), and the detection (bottom) for the HOM interference of the mode-engineered photons.
        The actual temporal modes (up to a global phase) of the photons are given at the bottom of each panel.
        To discern the network effect from the source imperfection, we check the HOM visibility of the source in each experiment that serves as the reference.
        The switching patterns and their timings for source quality checking are given on the right of the dashed line.
        (b) and (d): Plots show the respective HOM dips for the source, and the dotted line indicates the quantum-classical boundary. The visibilities obtained from the local coincidence measurements in both Figs. 2(a) and (b) from the main text very closely match with the corresponding source (reference) visibilities, thus highlighting the excellent network performance.
    }
    \label{fig:Switching_pattern_two-bin_tau0}
\end{figure*}

\begin{figure*}
    \centering
    \includegraphics[width=\textwidth]{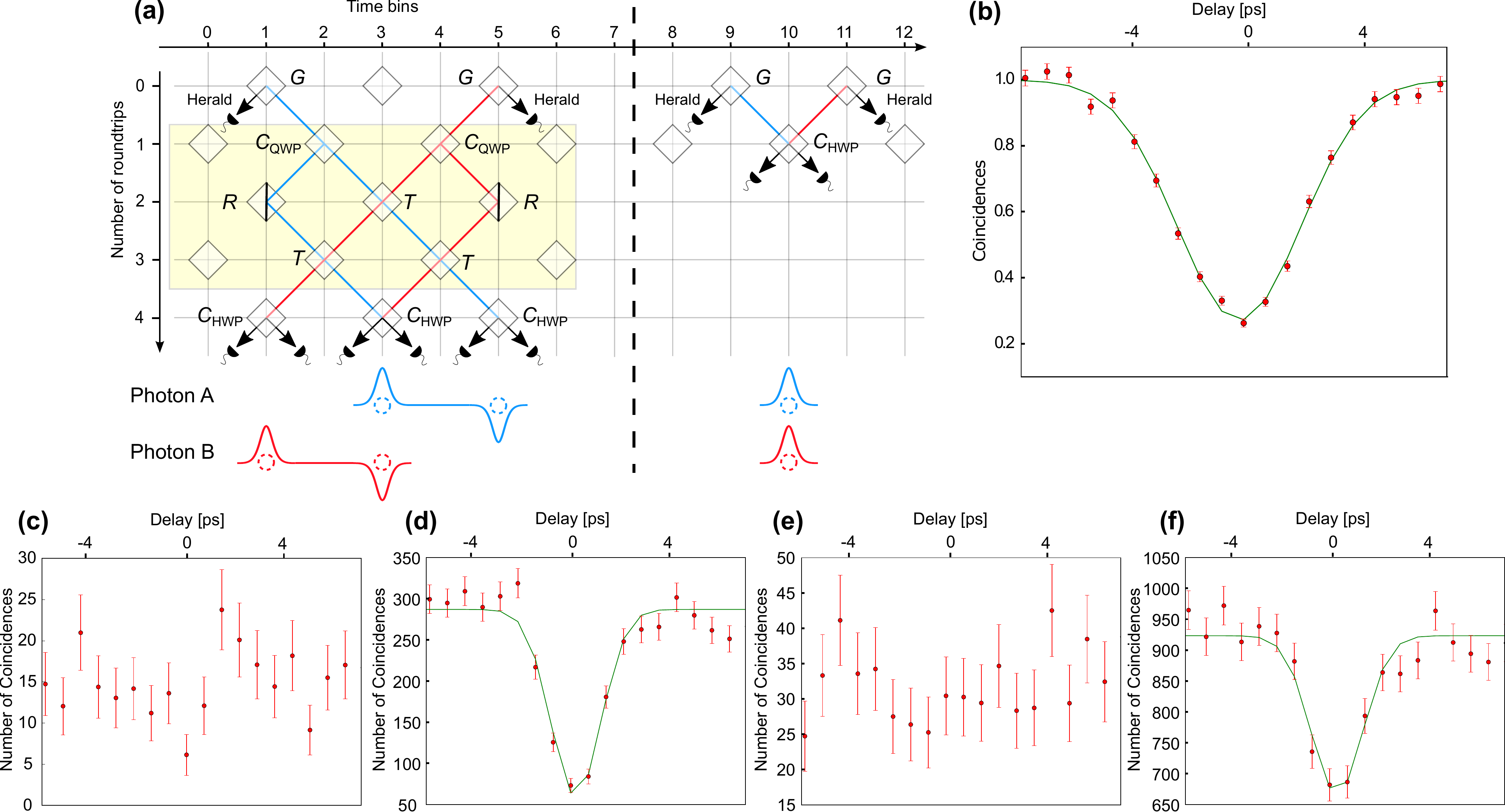}
    \caption{
        Here we include additional information to support the results presented in Figs. 2(c) and (d) of the main text.
        (a): The switching patterns implemented for the engineering of the photon wavepackets as well as for finding the source (reference) visibility.
        (b): HOM interference via coincidence for the source.
        (c)--(e): Results of the local measurements at the three different time bins, 1, 3, and 5 in (a).
        As expected, the number of coincidences at the two side bins stay very low (ideally zero) at all the delays since only one of the photon wavepackets is supposed to exist in these bins.
        (d): On the contrary, at the middle bin the two wavepackets coexist, and their interference leads to the HOM dip with a visibility approaching the source visibility.
        (f): Coincidence counts obtained from the global measurements.
        When the visibilities in (d) and (f) are normalized with the source visibility from (b), we obtain the results shown in  Fig. 2(c) and (d) from the main text, respectively.
        This again emphasizes the near-perfect performance of our network.
    } 
    \label{fig:Switching_pattern_two-bin_tau-1}
\end{figure*}

\begin{figure*}
    \centering
    \includegraphics[width=\textwidth]{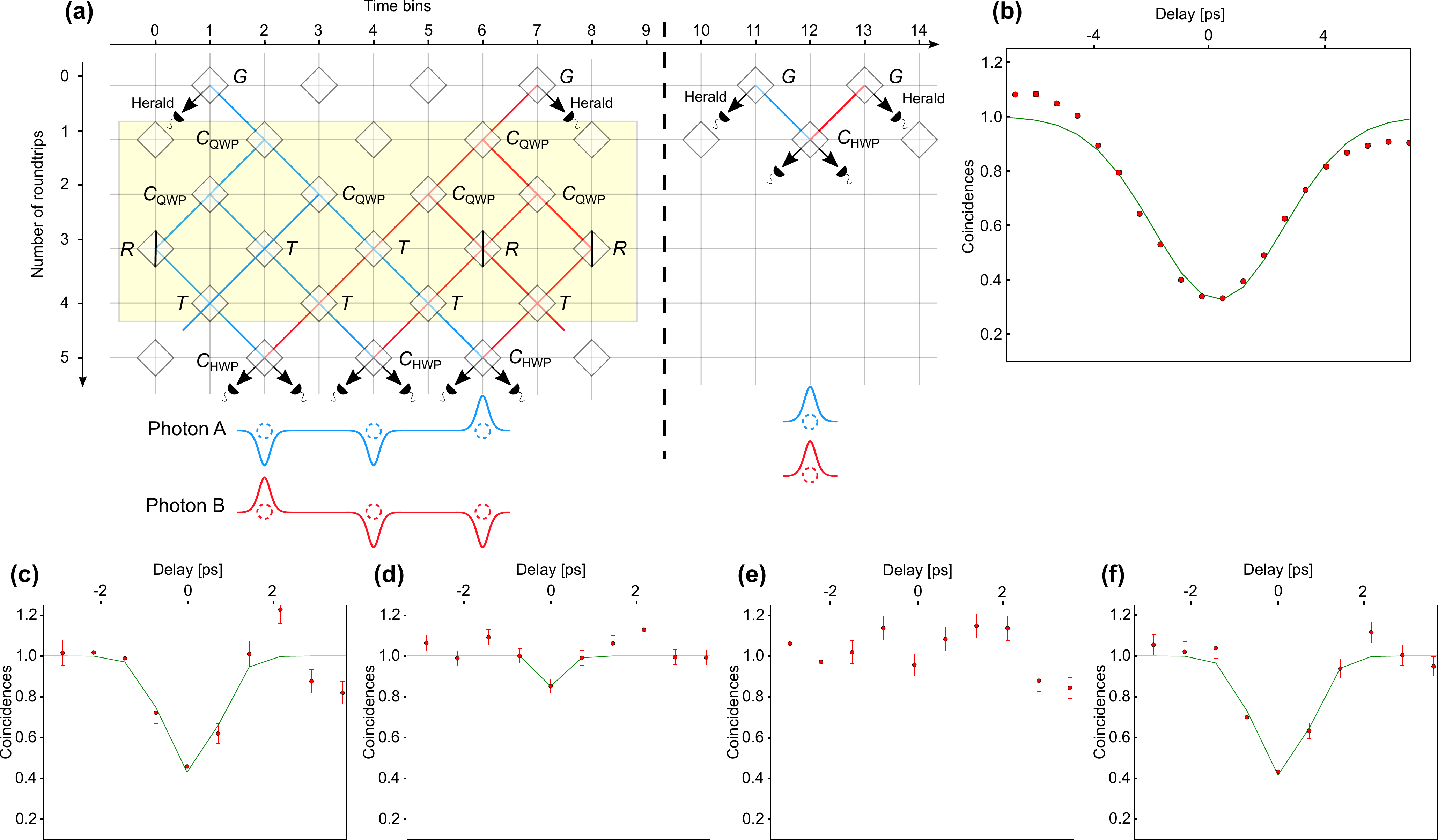}
    \caption{
        Supplementary information regarding the results of Figs. 2(e)--(h) of the main text is appended here.
        (a): The switching patterns implemented for synthesizing three-bin photon wavepackets and for checking the source (reference) visibility.
        (b): Coincidences for HOM interference of the source.
        (c)--(f): Measured coincidences for the measurement schemes presented in Fig. 2(e)--(h) of the main text, respectively.
        To discern the effect of the network on two-photon interference the values reported in Figs. 2(e)–(h) of the main text are obtained by normalizing the visibilities extracted from the data (c)–(f) by the source visibility extracted from (b).
        }  
    \label{fig:Switching_pattern_three-bin_tau0}
\end{figure*}


\begin{thebibliography}{99}
    \bibitem{Loudon73}
        R. Loudon, 
        \textit{The Quantum Theory of Light}
        (Oxford University Press, Oxford, UK, 1973).
    \bibitem{Donati73}
        O. Donati, G. P. Missiroli, and G. Pozzi,
        An experiment on electron interference,
        Am. J. Phys. \textbf{41}, 639 (1973). 
	\bibitem{HOM87}
		C. K. Hong, Z. Y. Ou, and L. Mandel,
		Measurement of Subpicosecond Time Intervals between Two Photons by Interference,
		Phys. Rev. Lett. \textbf{59}, 2044 (1987).
    \bibitem{Pan12}
        J.-W. Pan, Z.-B. Chen, C.-Y. Lu, H. Weinfurter, A. Zeilinger, and M \.{Z}ukowski,
        Multiphoton entanglement and interferometry,
        Rev. Mod. Phy. \textbf{84}, 777 (2012).
	\bibitem{KBMW98}
		A. Kuzmich , D. Branning , L. Mandel, and I. A. Walmsley,
		Multiphoton interference effects at a beamsplitter,
		J. Mod. Opt. \textbf{45}, 2233 (1998).
	\bibitem{Metal13}
		B. J. Metcalf \textit{et al.},
		Multiphoton quantum interference in a multiport integrated photonic device,
		Nat. Commun. \textbf{4}, 1356 (2013).
	\bibitem{TTSSGHNSW15}
		M. Tillmann, S.-H. Tan, S. E. Stoeckl, B. C. Sanders, H. de Guise, R. Heilmann, S. Nolte, A. Szameit, and P. Walther,
		Generalized Multiphoton Quantum Interference,
		Phys. Rev. X \textbf{5}, 041015 (2015).
	\bibitem{RFMWK16}
		L. Rigovacca, C. Di Franco, B. J. Metcalf, I. A. Walmsley, and M. S. Kim,
		Nonclassicality Criteria in Multiport Interferometry,
		Phys. Rev. Lett. \textbf{117}, 213602 (2016).
	\bibitem{AKJMSHRWJ17}
		S. Agne, T. Kauten, J. Jin, E. Meyer-Scott, J. Z. Salvail, D. R. Hamel, K. J. Resch, G. Weihs, and T. Jennewein,
		Observation of Genuine Three-Photon Interference,
		Phys. Rev. Lett. \textbf{118}, 153602 (2017).
	\bibitem{NWXC18}
		\'A. Navarrete, W. Wang, F. Xu, and M. Curty,
		Characterizing multiphoton quantum interference with practical light sources and threshold single-photon detectors,
		New J. Phys. \textbf{20}, 043018 (2018).
    \bibitem{Wiseman10}
        H. M. Wiseman and G. J. Milburm,
        \textit{Quantum Measurement and Control}
        (Cambridge University Press, Cambridge, England, 2010). 
	\bibitem{CGW13}
		A. M. Childs, D. Gosset, and Z. Webb,
		Universal computation by multiparticle quantum walk,
		Science \textbf{339}, 791 (2013).
	\bibitem{KLM01}
		E. Knill, R. Laflamme, and G. J. Milburn,
		A scheme for efficient quantum computation with linear optics,
		Nature (London) \textbf{409}, 46 (2001).
	\bibitem{Raussendorf01}
	    R. Raussendorf, and H. J. Briegel, 
	    A One-Way Computer,
	    Phys. Rev. Lett. \textbf{86}, 5188 (2001). 
	\bibitem{O'brien09}
    	J. L. O'Brien, A. Furusawa, and J. Vu\v{c}kovi\'{c}
    	Photonic quantum technologies,
    	Nat. Photonics \textbf{3}, 687 (2009). 
	\bibitem{SA88}
		Y. H. Shih and C. O. Alley,
		New Type of Einstein-Podolsky-Rosen-Bohm Experiment Using Pairs of Light Quanta Produced by Optical Parametric Down Conversion,
		Phys. Rev. Lett. \textbf{61}, 2921 (1988).
	\bibitem{KSC92}
	    P. G. Kwiat, A. M. Steinberg, and R. Y. Chiao,
	    Observation of a ‘‘quantum eraser’’: A revival of coherence in a two-photon interference experiment,
        Phys. Rev. A \textbf{45}, 7729 (1992).
	\bibitem{LWHRK04}
		T. Legero, T. Wilk, M. Hennrich, G. Rempe, and A. Kuhn,
		Quantum Beat of Two Single Photons,
		Phys. Rev. Lett. \textbf{93}, 070503 (2004).
    \bibitem{MJMTBKW17}
        A. J. Menssen, A. E. Jones, B. J. Metcalf, M. C. Tichy, S. Barz, W. S. Kolthammer, and I. A. Walmsley,
        Distinguishability and Many-Particle Interference,
        Phys. Rev. Lett. \textbf{118}, 153603 (2017).
	\bibitem{RSG19}
		M. Rezai, J. Sperling, and I. Gerhardt,
		What can single photons do what lasers cannot do?,
		Quantum Sci. Technol. \textbf{4}, 045008 (2019).
	\bibitem{ADZ93}
		Y. Aharonov, L. Davidovich, and N. Zagury,
		Quantum random walks,
		Phys. Rev. A \textbf{48}, 1687 (1993).
	\bibitem{SCPGMAJS10}
		A. Schreiber, K. N. Cassemiro, V. Poto\v{c}ek, A. G\'{a}bris, P. J. Mosley, E. Andersson, I. Jex, and C. Silberhorn,
		Photons Walking the Line: A Quantum Walk with Adjustable Coin Operations,
		Phys. Rev. Lett. \textbf{104}, 050502 (2010).
	\bibitem{NBKSSGPKJS18}
		T. Nitsche, S. Barkhofen, R. Kruse, L. Sansoni, M. \v{S}tefa\v{n}\'{a}k, A. G\'{a}bris, V. Poto\v{c}ek, T. Kiss, I. Jex, and C. Silberhorn,
		Probing measurement-induced effects in quantum walks via recurrence,
		Sci. Adv. \textbf{4}, eaar6444 (2018).
	\bibitem{AA11}
		S. Aaronson and A. Arkhipov,
		The computational complexity of linear optics,
		in Proceedings of the Forty-Third Annual ACM Symposium on Theory of Computing, (ACM), STOC 11
		(Association for Computing Machinery, New York, 2011), pp. 333–342.
	\bibitem{MGDR14}
		K. R. Motes, A. Gilchrist, J. P. Dowling, and P. P. Rohde,
		Scalable Boson Sampling with Time-Bin Encoding Using a Loop-Based Architecture,
		Phys. Rev. Lett. \textbf{113}, 120501 (2014).
	\bibitem{Hetal17}
		Y. He \textit{et al.},
		Time-Bin-Encoded Boson Sampling with a Single-Photon Device,
		Phys. Rev. Lett. \textbf{118}, 190501 (2017).
	\bibitem{W20}
	    M. Walschaers,
	    Signatures of many-particle interference,
	    J. Phys. B: At. Mol. Opt. Phys. \textbf{53}, 043001 (2020).
	\bibitem{MW95}
		L. Mandel and E. Wolf,
		\textit{Optical Coherence and Quantum Optics}
		(Cambridge University Press, Cambridge, England, 1995).
	\bibitem{Nitsche16}
	    T. Nitsche, F. Elster, J. Novotn\'{y}, A. G\'{a}bris, I. Jex, S. Barkhofen, and C. Silberhorn, 
	    Quantum walks with dynamical control: Graph engineering, initial state preparation and state transfer,
	    New J. Phys. \textbf{18}, 063017 (2016).
	\bibitem{SM}
		See Supplemental Material below for details on the experiment and additional results.
	\bibitem{HABDMS13}
		G. Harder, V. Ansari, B. Brecht, T. Dirmeier, C. Marquardt, and C. Silberhorn,
		An optimized photon pair source for quantum circuits,
		Opt. Express \textbf{21}, 13975 (2013).
	\bibitem{KDM77}
	    H. J. Kimble, M. Dagenais, and L. Mandel,
	    Photon Antibunching in Resonance Fluorescence,
	    Phys. Rev. Lett. \textbf{39}, 691 (1977).
	\bibitem{Guzik12}
	    A. Aspuru-Guzik and P. Walther, 
	    Photonic quantum simulators,
	    Nat. Phys. \textbf{8}, 285 (2012).
    \bibitem{Bennett2005}
        C. H. Bennett, P. Hayden, D. W. Leung, P. W. Shor, and A. Winter, 
        Remote preparation of quantum states,
        IEEE Trans. Inf. Theory \textbf{51}, 56 (2005).
\end{thebibliography}
\end{document}